\documentclass[conference,a4paper,twocolumn]{IEEEtran}

\IEEEoverridecommandlockouts

\ifCLASSINFOpdf
\else
\fi
  \usepackage{tipa}
  \usepackage{graphicx}
	\usepackage{epsfig}
	\usepackage{epstopdf}
	\usepackage{upgreek}
	\usepackage[nolist,nohyperlinks]{acronym}

\usepackage[cmex10]{amsmath}
\usepackage{subfigure}
\newcommand{\noindento}{\vspace{0.1cm} \noindent}

\begin{acronym}
\acro{WNoC}{Wireless Network-on-Chip}
\acro{SDM}{Software-Defined Metamaterial}
\acro{HEMT}{High Electron Mobility Transistor}
\acro{SPP}{Surface Plasmon Polariton}
\acro{LIM}{Low Index Material}
\acro{HIM}{High Index Material}
\end{acronym}

\begin{document}



\title{Graphene-based Terahertz Antennas for Area-Constrained Applications}


\author{
\IEEEauthorblockN{
Sergi Abadal, Seyed Ehsan Hosseininejad, Albert Cabellos-Aparicio, Eduard Alarc\'{o}n
}
\IEEEauthorblockA{
NaNoNetworking Center in Catalunya (N3Cat)\\
Universitat Polit\`{e}cnica de Catalunya (UPC), Barcelona, Spain
}
\thanks{The authors gratefully acknowledge support from the {\em Spanish Ministry of Economy and Competitiveness} under grant PCIN-2015-012 and from the EU's Horizon 2020 research and innovation programme under grant agreement No. 736876.} \thanks{Corresponding email: abadal@ac.upc.edu}
}

\maketitle


\begin{abstract}
Graphene is enabling a plethora of applications in a wide range of fields due to its unique electrical, mechanical, and optical properties. In the realm of wireless communications, graphene shows great promise for the implementation of miniaturized and tunable antennas in the terahertz band. These unique advantages open the door to disruptive wireless applications in highly integrated scenarios where conventional communications means cannot be employed. In this paper, recent advances in plasmonic graphene antennas are presented. \ac{WNoC} and \acp{SDM}, two new area-constrained applications uniquely suited to the characteristics of graphene antennas, are then described. The challenges in terms of antenna design and channel characterization are outlined for both case scenarios.
\end{abstract}


\begin{IEEEkeywords}
Area-Constrained Applications; Graphene Antennas; Miniaturized Antennas; Plasmonic Antennas; Short-range Communications; Terahertz Band; Tunable Antennas
\end{IEEEkeywords}

%
\IEEEpeerreviewmaketitle




\acresetall

\section{Introduction}
Terahertz technology has made significant advances in the fields of spectroscopy, imaging and, more recently, wireless communications. In the latter, the use of this frequency band between 0.1 and 10 THz becomes extremely attractive due to the abundance of bandwidth and the potential for low area and power footprints, yet challenging given the large propagation losses and the lack of mature devices and circuits for terahertz operation. Maturity issues aside, this combination of features renders terahertz wireless communications desirable for highly integrated applications where area may be a decisive metric.

In the pursuit of fast and efficient means to transmit terahertz signals, graphene appears an excellent candidate for the implementation of new devices. Due to its unique electrical and optical properties, the potential of graphene has been indeed evaluated in the context of analog RF transistors~\cite{Schwierz2013}, metamaterials \cite{Vakil2011}, or broadband nanophotonics \cite{Bao2012}, among others. Many of these works have highlighted the capacity of graphene to support \ac{SPP} waves in the terahertz regime, which results in extreme confinement or tunability. This has allowed researchers to create fast and compact devices with unprecedented reconfigurability capabilities in the terahertz band.

\begin{figure}[!t]
\centering
\includegraphics[width=2.8in]{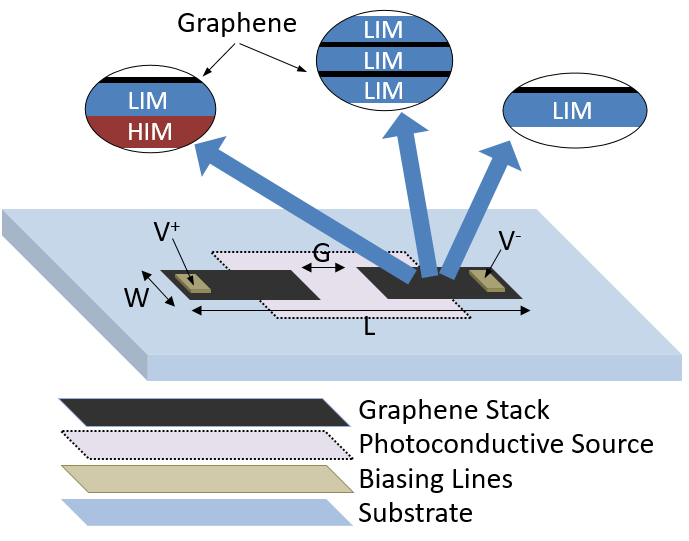}
\vspace{-0.25cm}
\caption{Schematic representation of a $W\times L$ graphene-based dipole with a gap of length $G$ and fed with a photoconductive source. The radiating elements can take the form of different graphene-based stacks, each with its own advantages.}
\vspace{-0.3cm}
\label{fig:graphenna}\label{fig:antenna}
\end{figure} 

In this paper, we focus on the application of graphene in the field of antennas \cite{AbadalTCOM}. We first review current practices by analyzing designs from the literature that use graphene either as the radiating element (see Fig.~\ref{fig:graphenna}) or in other parts of the antenna, e.g. the feed. Through the analysis, we discuss the miniaturization and reconfigurability opportunities enabled by graphene in these devices. Finally, we describe two new disruptive applications that will greatly benefit from the unique properties of graphene-based antennas: \acp{WNoC} and \acp{SDM}. As we will see, both applications share the need for smaller form factors of wireless communication while maintaining certain application-dependent performance requirements.

The remainder of the paper is as follows. Section~\ref{sec:antennas} outlines the fundamentals of graphene antennas, to then review a set of existing designs. Section~\ref{sec:apps} provides a context analysis of applications where graphene antennas can play a key role.
Finally, Section~\ref{sec:conclusion} concludes the paper. 



\section{Graphene-based Terahertz Antennas}
\label{sec:antennas}
Graphene, a flat monolayer of carbon atoms tightly packed in a two-dimensional honeycomb lattice, has recently attracted the attention of the research community due to its extraordinary 
mechanical, electronic, and optical properties~\cite{Geim2007}. In antenna theory, first works followed findings by Hanson {\em et al.} on the propagation of electromagnetic waves on laterally-infinite graphene layers \cite{Hanson2008a}. We review the formulations used to model the conductivity of graphene sheets in Section~\ref{sec:conductivity}.

The plasmonic nature of the graphene conductivity in the terahertz band, where metals act as lossy non-plasmonic conductors, enables the creation of miniaturized and tunable devices. As we will see in Section~\ref{sec:radiating}, these are the main arguments in favor of the use of graphene as the radiating element of terahertz antennas. First proposals, however, show a rather low efficiency and suggest that leveraging the tunability of graphene in elements other than the radiator would be more appropriate \cite{Perruisseau-Carrier2013}. These are discussed in Section~\ref{sec:auxiliary}.

\subsection{Conductivity Model}
\label{sec:conductivity}
\begin{figure}[!t]
\centering
\subfigure[$\tau$ = 1 ps\label{fig:condEf}]{\subfigtopskip = 0pt \subfigbottomskip = -15pt \includegraphics[width=1.7in]{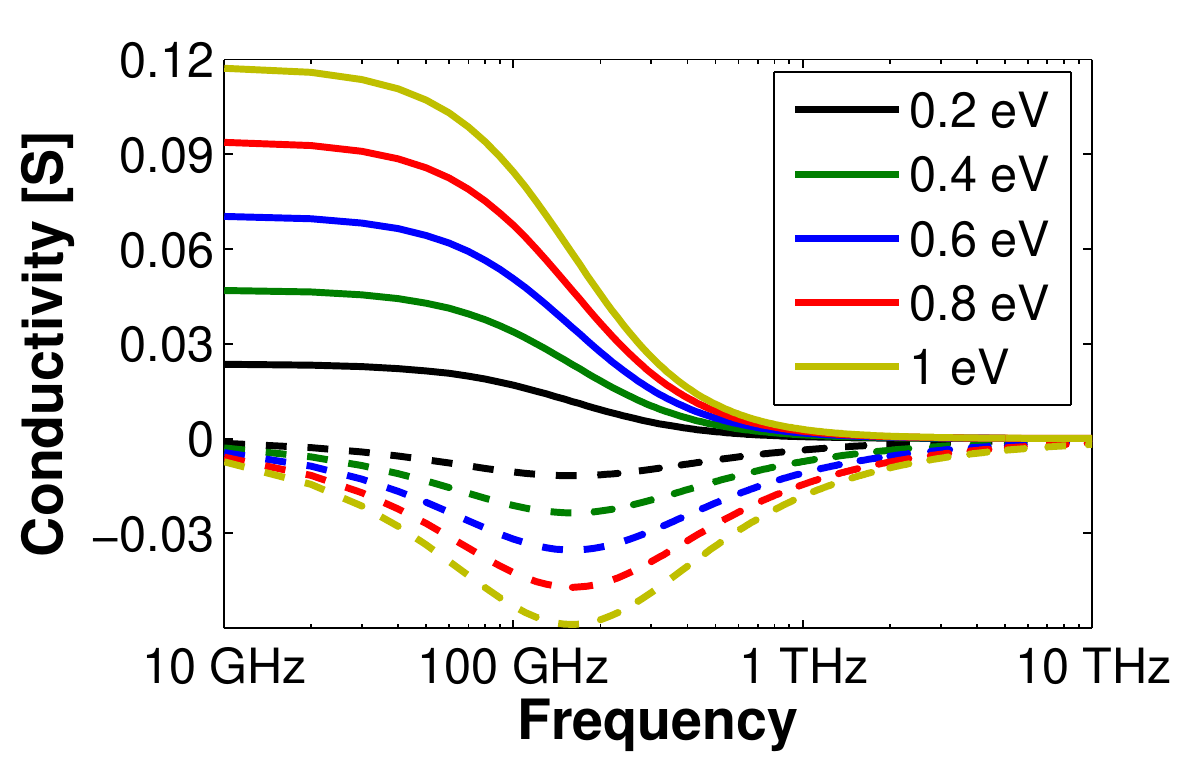}}
\subfigure[$E_{F}$ = 0.6 eV\label{fig:condTau}]{\subfigtopskip = 0pt \subfigbottomskip = -15pt \includegraphics[width=1.7in]{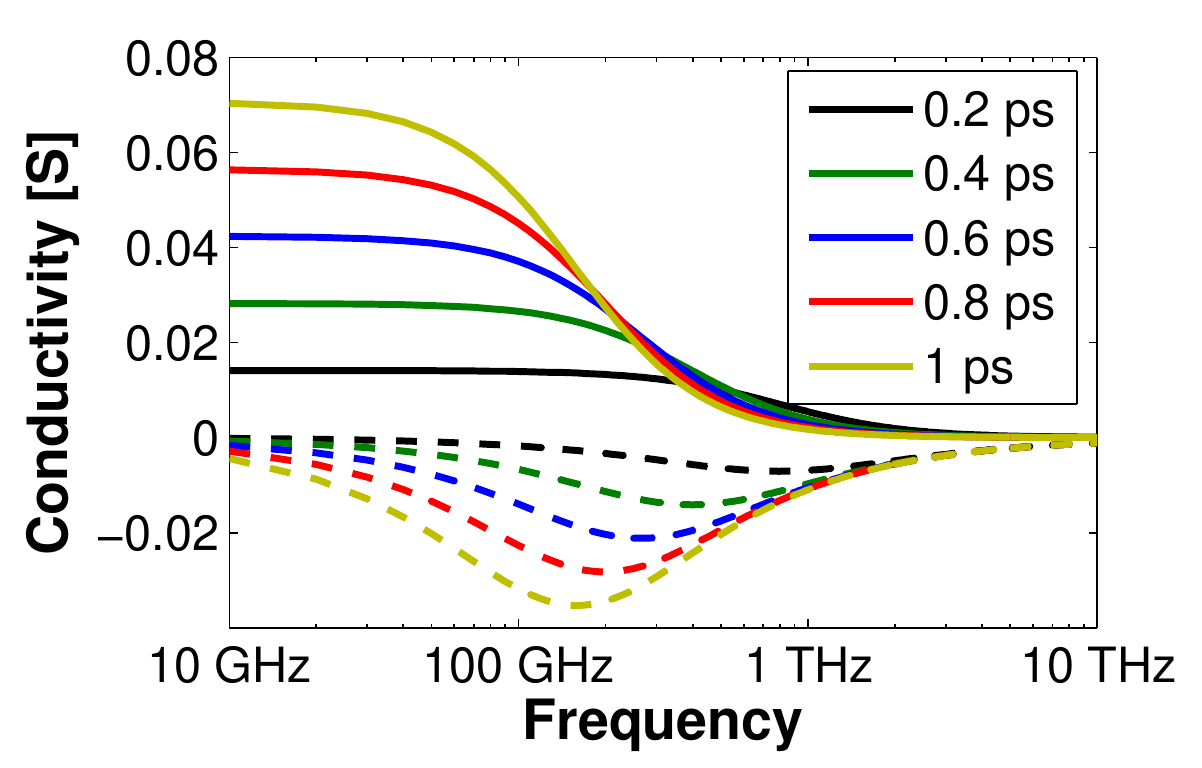}}
\vspace{-0.1cm}
\caption{Graphene conductivity for different chemical potential and relaxation time values. Solid and dotted lines represent $\Re{[\sigma]}$ and $-\Im{[\sigma]}$, respectively.}
\label{fig:conductivity}
\vspace{-0.3cm}
\end{figure}

The Kubo formalism is generally used to model the conductivity of graphene at terahertz frequencies \cite{Hanson2008a}. The main approach considers that the sheets used in graphene-based antennas are large enough to disregard the effects of the graphene edges. Moreover, experimental results show that the Drude-like intraband contribution dominates, so that the conductivity can be expressed as
\begin{equation}
\sigma\left(\omega\right)=\frac{2e^{2}}{\pi\hbar}\frac{k_{B}T}{\hbar}\ln\left[2\cosh\left[\frac{E_{F}}{2k_{B}T}\right]\right]\frac{i}{\omega+i\tau^{-1}},\label{eq:sigma_graphene}
\end{equation}
where $e$, $\hbar$, and $k_{B}$ are constants corresponding to the charge of an electron, the reduced Planck constant, and the Boltzmann constant, respectively. Variables $T$, $E_{F}$, and $\tau$ correspond to the temperature, the chemical potential, and the relaxation time of the graphene layer. Figure~\ref{fig:conductivity} plots the conductivity as a function of the latter two variables, which have a strong impact on the resonant frequency and radiation efficiency of the antennas. With $\sigma(\omega)$, graphene can be modeled as an infinitesimally thin surface with impedance $Z(\omega) = \sigma^{1}(\omega)$.

The chemical potential $E_{F}$ refers to the level in the distribution of electron energies at which a quantum state is equally likely to be occupied or empty. The value of the chemical potential can be changed by means of an electrostatic bias or chemical doping, therefore providing means for reconfigurability. In general, the chemical potential variation $\Delta E_{F}$ relates to the variation in voltage $\Delta V$ as $\Delta E_{F} \propto \sqrt{|\Delta V|}$ \cite{Yu2009Chemical}. As implied Fig.~\ref{fig:condEf}, higher chemical potential leads to better performance through an increase of the conductivity.


%

The relaxation time $\tau$ is the interval required for a material to restore a uniform charge density after a charge distortion is introduced. Its value mainly depends on the interactions of electrons with scatterers, defects, and other phenomena that depend on the quality of the graphene sample, among others. In summary, sheets of high quality graphene will exhibit a large relaxation time which, as implied in Fig.~\ref{fig:condTau}, is desirable to increase the conductivity and provide good antenna performance. 

\subsection{Graphene as the Radiating Element}
\label{sec:radiating}
\begin{figure}[!t]
\centering
\subfigure[$\Re{[Z]}$ (solid) and $\Im{[Z]}$ (dotted).\label{fig:L1}]{\subfigtopskip = 0pt \subfigbottomskip = -15pt \includegraphics[width=2.25in]{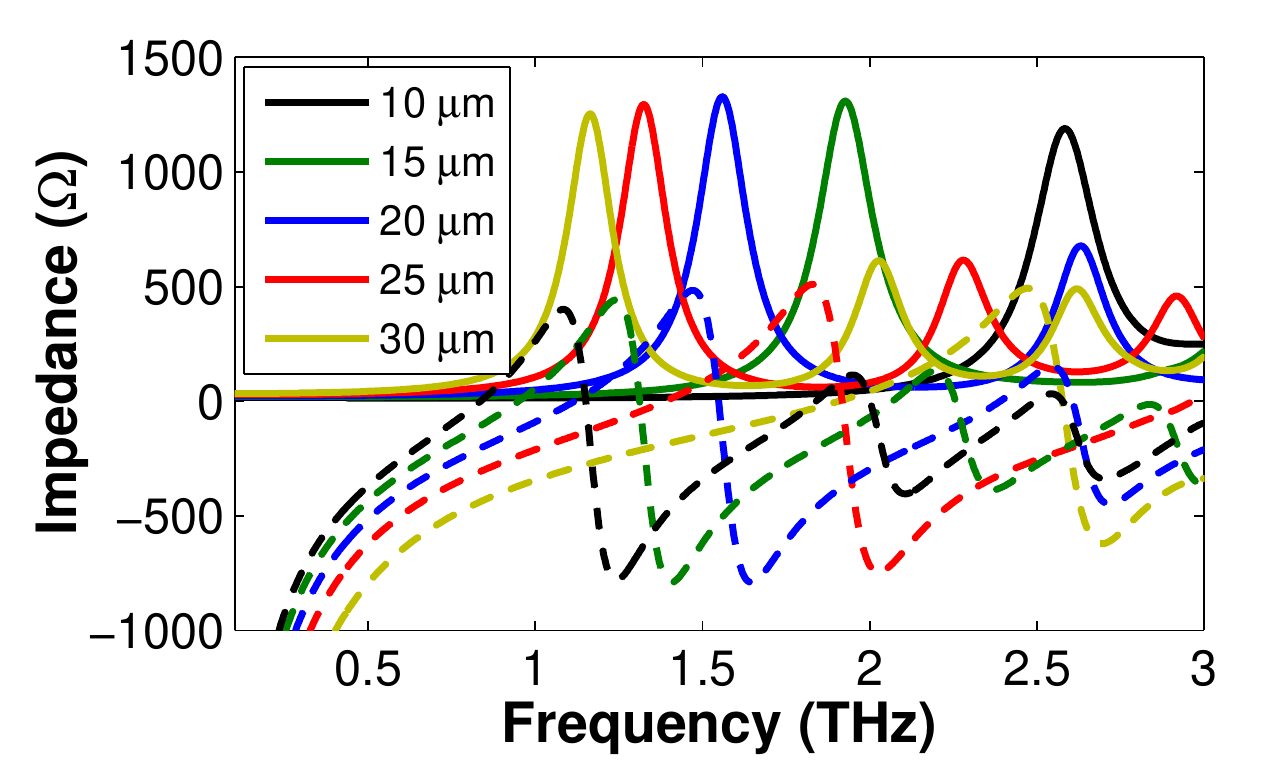}}
\subfigure[Resonance point.\label{fig:L2}]{\subfigtopskip = 0pt \subfigbottomskip = -15pt \includegraphics[width=1.1in]{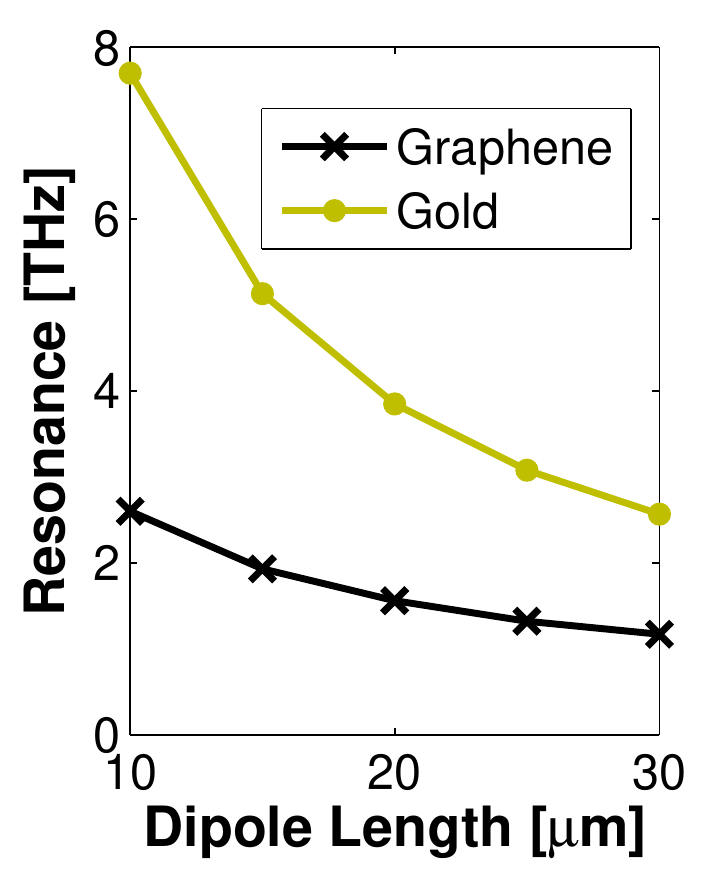}}
\vspace{-0.1cm}
\caption{Resonance characteristics of the graphene dipole as a function of the length $L$.}
\vspace{-0.3cm}
\label{fig:L}
\end{figure}

First works on propagation of terahertz waves over graphene sheets led to a surge of proposals that, in essence, consist of a number of finite-size graphene layers (the radiating elements) mounted over a dielectric material and a feed to drive the signals to the antenna \cite{AbadalTCOM}. These antennas have been studied from different perspectives, namely:

\noindento {\bfseries Antenna dimensions:} We use CST MWS \cite{CST} to simulate the dipole shown in Fig.~\ref{fig:antenna} with a simple stack of graphene over \ac{LIM}. The width is $W$ = 8 $\upmu$m, the gap is $G$ = 3 $\upmu$m, the substrate is quartz with permittivity $\varepsilon_{r}$ = 3.8, whereas $\tau$ = 1 ps and $E_{F}$ = 0.2 eV. The length $L$ is variable. Fig.~\ref{fig:L1} shows how longer dipoles lead to lower resonance points, as one would expect. Note, however, that metallic dipoles of the same length would theoretically resonate at much higher frequencies, as observed in Fig.~\ref{fig:L2} (we took the working point where $\Im{[Z]}=0$ and $\Re{[Z]}$ is high). The use of graphene does not change the radiation pattern at resonance as proved in other works \cite{AbadalTCOM}. The maximum gain (IEEE) and the simulated radiation efficiency are relatively low in the considered cases, around -10 dB and 4.5\%, respectively. 

\begin{figure}[!t]
\centering
\subfigure[$\Re{[Z]}$ (solid) and $\Im{[Z]}$ (dotted).\label{fig:chem1}]{\subfigtopskip = 0pt \subfigbottomskip = -15pt \includegraphics[width=2.1in]{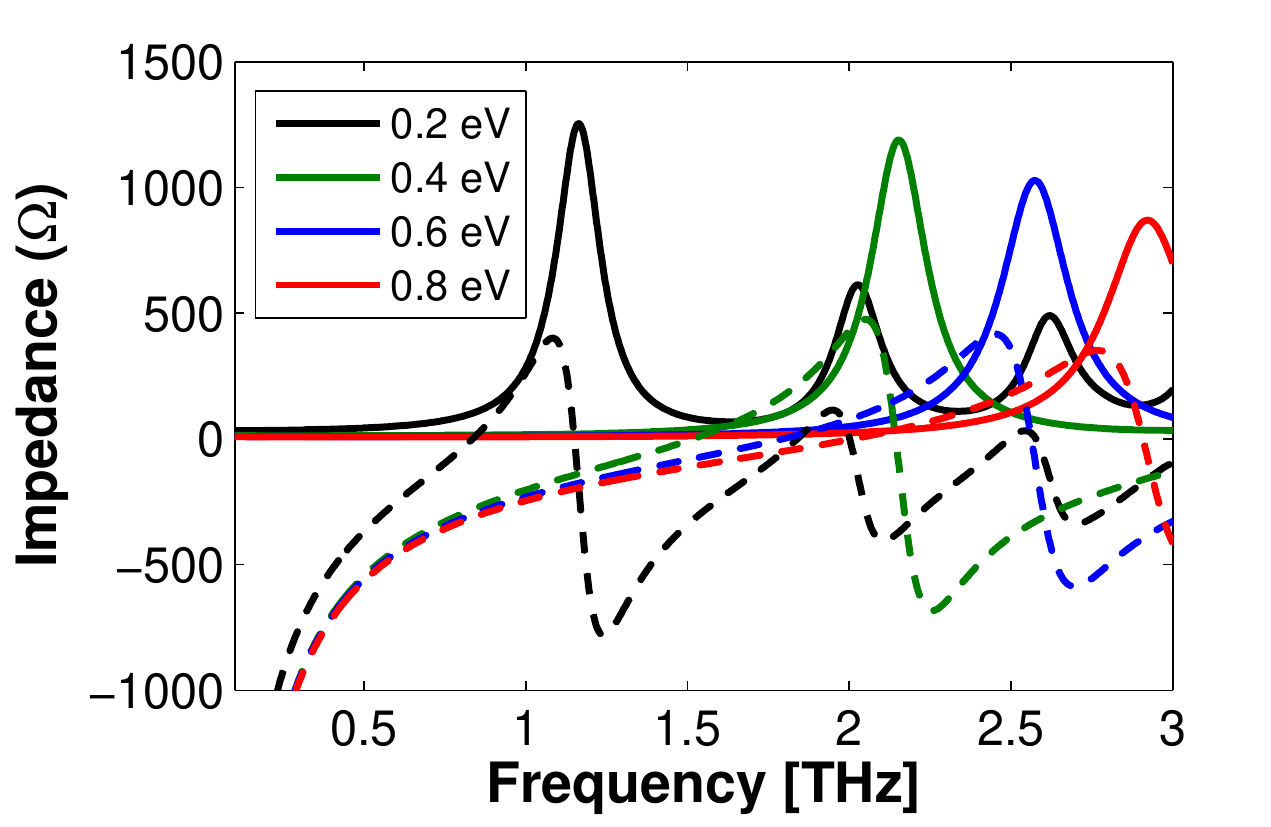}}
\subfigure[Resonance / efficiency.\label{fig:chem2}]{\subfigtopskip = 0pt \subfigbottomskip = -15pt \includegraphics[width=1.3in]{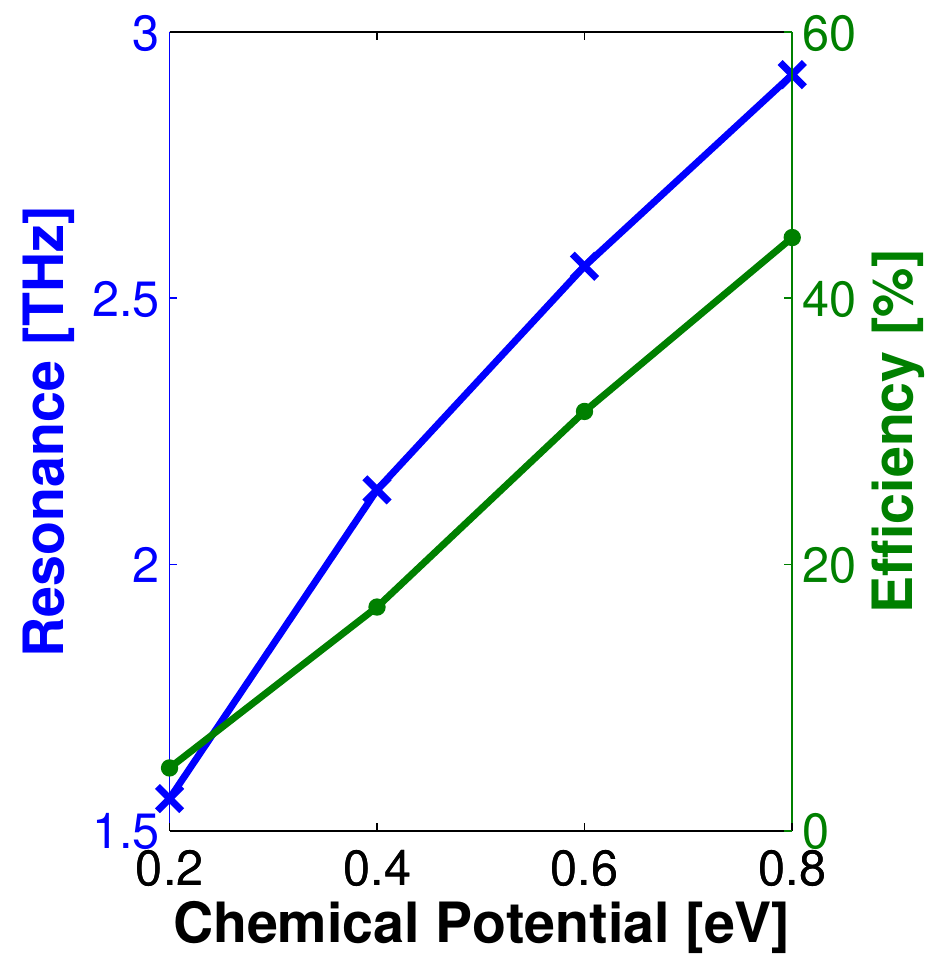}}
\vspace{-0.3cm}
\caption{Resonance characteristics of the graphene dipole as a function of the chemical potential $E_{F}$.}
\vspace{-0.3cm}
\label{fig:chem}
\end{figure}

\noindento {\bfseries Chemical potential:} Consider a dipole with $W$ = 8 $\upmu$m, $L$ = 20 $\upmu$m, $G$ = 3 $\upmu$m, with $\tau$ = 1 ps and $\varepsilon_{r}$ = 3.8. Fig.~\ref{fig:chem} shows the resonance characteristics of the antenna as a function of the chemical potential. The increase of chemical potential leads to a significant shift of the resonant frequency and to an enhancement of the antenna response without changes in the radiation pattern. The efficiency and gain increase (52\% and -0.46 dB for $E_{F}$ = 0.8 eV), but at the expense of having to apply an increasing bias. 

\begin{figure}[!t]
\centering
\subfigure[$\Re{[Z]}$ (solid) and $\Im{[Z]}$ (dotted).\label{fig:relax1}]{\subfigtopskip = 0pt \subfigbottomskip = -15pt \includegraphics[width=2.15in]{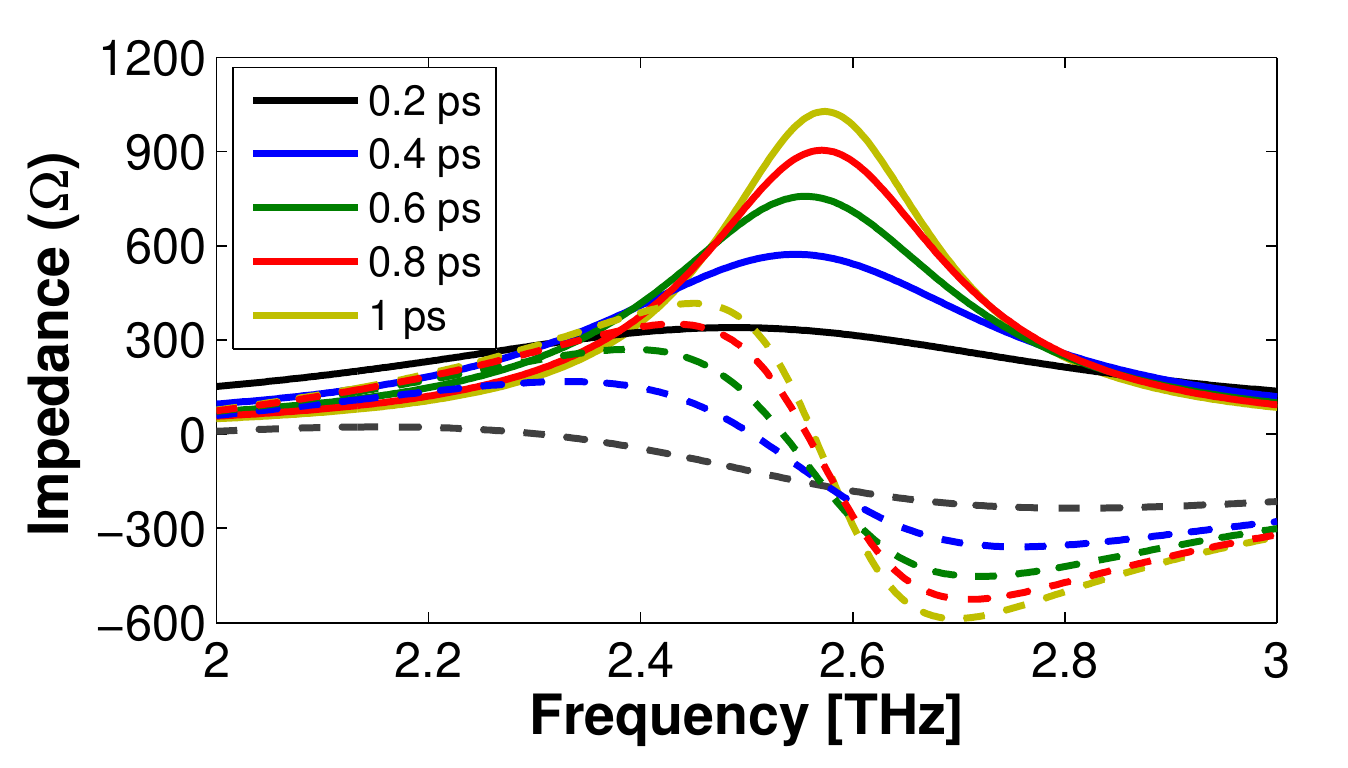}}
\subfigure[Efficiency.\label{fig:relax2}]{\subfigtopskip = 0pt \subfigbottomskip = -15pt \includegraphics[width=1.25in]{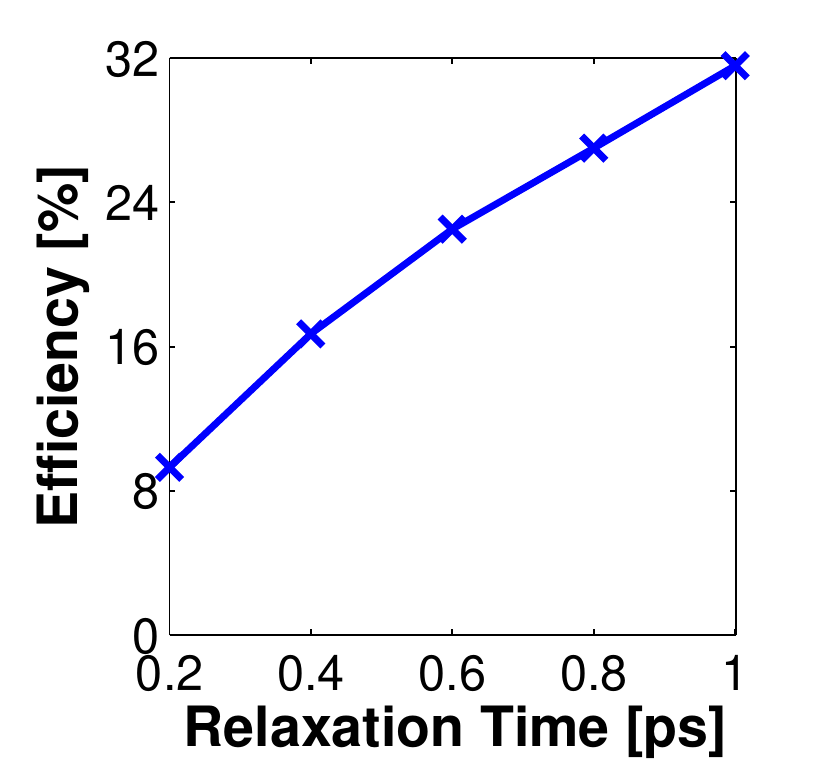}}
\vspace{-0.1cm}
\caption{Resonance characteristics of the graphene dipole as a function of the relaxation time $\tau$.}
\vspace{-0.3cm}
\label{fig:relax}
\end{figure}

\noindento {\bfseries Relaxation time:} Now consider a dipole with $E_{F}$ = 0.6 eV and variable $\tau$. Fig.~\ref{fig:relax} shows how the antenna maintains the same resonant frequency as the relaxation time is increased, but exhibiting a stronger resonant behavior. This translates into a higher efficiency and gain (32\% and -2.34 dB for $\tau$ = 1 ps) due to the better carrier mobility of the graphene sheets. Achieving such high relaxation times, however, requires the use of substrates such as boron nitride \cite{Cabellos2014} or advances in the fabrication of monolayers. 

\noindento {\bfseries Radiating element stack:} although laying monolayer graphene sheets to form an antenna produces interesting results, more complex stacks like that depicted in Fig.~\ref{fig:antenna} have been examined. G\'{o}mez-D\'{i}az {\em et al.} propose to place two graphene monolayers separated by a thin dielectric to achieve biasing without the need of additional gates \cite{Gomez2015}. 

\begin{figure}[!t]
\centering
\includegraphics[width=2.5in]{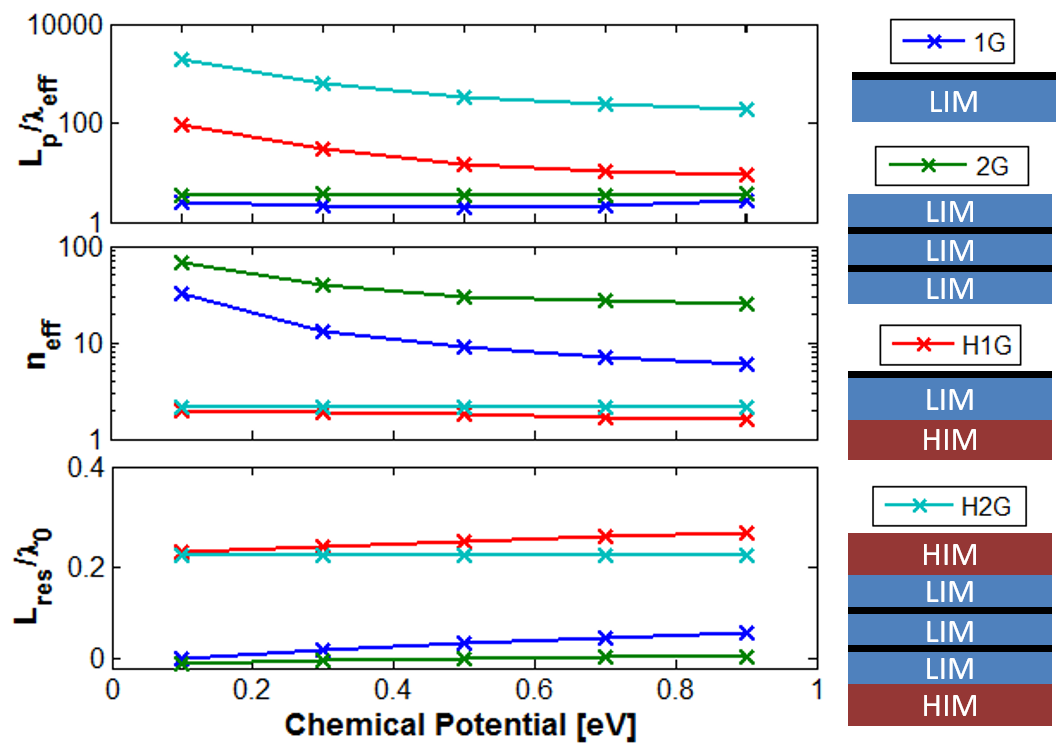}
\vspace{-0.2cm}
\caption{Different stack configurations seen from the perspective of efficiency (top), tunability (center), and miniaturization (bottom).}
\vspace{-0.4cm}
\label{fig:stacks}
\end{figure}

Also, our previous work in \cite{Hosseininejad2017} considers multiple stack combinations of graphene sheets with \ac{LIM} and \ac{HIM} layers to increase the efficiency of the antenna for fixed $\tau$ and $E_{F}$. To evaluate this, we simulated the different stacks shown in the right part of Fig.~\ref{fig:stacks} for different $E_{F}$ values and considering $\tau = 0.6$ ps and the dimensions indicated in \cite{Hosseininejad2017}. The top chart of Fig.~\ref{fig:stacks} shows how the efficiency, measured in terms of the normalized propagation length $L_{p}$ of the \ac{SPP} waves within the stack, is significantly improved in hybrid configurations, i.e. those including \acp{HIM}. Such increase is achieved by means of the coupling of plasmons with dielectric waveguide modes, but at the expense of a reduction of the tunability and miniaturization. The loss of tunability is demonstrated by the small variation in terms of effective refractive index $n_{eff}$ observed in the center plot of Fig.~\ref{fig:stacks} for the H1G and H2G stacks. The loss of miniaturization is shown in the bottom plot: hybrid configurations have a much larger resonant length $L_{res}$. These results uncover interesting tradeoffs for the design of graphene-based terahertz antennas. We refer the interested reader to \cite{Hosseininejad2017} for more details.

\noindento {\bfseries Terahertz source:} Several works have discussed the possibility of employing a photomixer to approach the antenna. Its impedance, typically on the order of 10 $k\Omega$, leads to a reasonable matching with the relatively high impedance of the graphene dipole \cite{Perruisseau-Carrier2013}. Similarly, Cabellos {\em et al.} have proposed to use a photoconductor working in pulsed mode to excite the antenna \cite{Cabellos2014}. In this configuration, a photoconductive material (e.g. LT-GaAs) would be placed at the gap between the dipole arms and would be illuminated with a laser with femtosecond pulses. If an electrostatic bias is applied, this generates a photocurrent with terahertz components that enters the antenna. A unique trait of graphene photoconductive antennas is that the bias voltage could have a double effect on performance since increasing the bias theoretically augments both the generated photocurrent and the antenna efficiency. 

Although the use of photoconductive sources may be useful for the experimental validation of graphene antennas, it results impractical for area-constrained applications. Instead, compact electronic terahertz sources are required. In this direction, Jornet {\em et al.} propose the use of a \ac{HEMT} with graphene-based gate. If the transistor is sized properly, the application of voltage generates a terahertz plasma wave at the gate \cite{Jornet2014TRANSCEIVER}. This wave can be fed into the graphene antenna without contact issues and reducing the impedance mismatch problems. However, a direct comparison of the efficiency of \ac{HEMT}-based and photoconductive sources becomes challenging due to their radically different working principles. 

\subsection{Graphene as an Auxiliary Element} \label{sec:auxiliary}
Another research line attempts to exploit the reconfigurability potential of graphene and to avoid losses by not using it as the radiating element. For instance, a few works propose to use graphene sheets to provide frequency tunability to a metallic structure that can radiate with a fair efficiency in the terahertz band \cite{Hosseininejad2016a}. Also, in \cite{Huang2012ARRAY}, the authors implement the ground plane of a terahertz antenna using an array of graphene patches that form a switchable High Impedance Surface (HIS), such that by reconfiguring the HIS at real time, beam steering can be performed in a fast and fine-grained fashion. Others have proposed to use a similar technique for the design of reflectarrays with very low complexity \cite{Carrasco2013}.

\section{Area-Constrained Applications}
\label{sec:apps}
Wireless NanoSensor Networks (WNSNs) are one of the first wireless applications with evident area constraints suggesting the use of microscale antennas, as nanosensors are expected to have lateral dimensions of a few tens of microns. For this application, graphene-based terahertz antennas have been already considered as key enablers because of their miniaturization properties \cite{Akyildiz2010}.  

Here, we go beyond WNSNs and describe two new area-constrained applications that are uniquely suited to the characteristics of graphene-based terahertz antennas. Their main communication requirements are summarized in Table~\ref{tab:reqs}.



\begin{table}[!t] 
\caption{Area-Constrained Applications Characteristics}
\vspace{-0.1cm}
\label{tab:reqs}
\footnotesize
\centering
\begin{tabular}{lccc} 
\hline
 & {\bf WNSN \cite{Akyildiz2010}} & {\bf SDM \cite{AbadalACCESS}} & {\bf WNoC \cite{AbadalMICRO}} \\
\hline
Node Size & 1--100 $\upmu$m\textsuperscript{2} & 0.01--100 mm\textsuperscript{2} & 0.01--1 mm\textsuperscript{2}\\
TX Range & 0.1--100 cm  & 0.1--100 cm  &  0.1--10 cm \\
Data Rate & 0.001--0.1 Gbps & 0.01--1 Gbps & 10--100 Gbps \\
\hline
\end{tabular}
\vspace{-0.5cm}
\end{table}

\noindento {\bf Software-Defined Metamaterials (SDM).} Metamaterials have recently enabled the realization of novel electromagnetic components with engineered functionalities, but have very limited room for reusability or adaptability. The \ac{SDM} paradigm aims to address this issue by providing reconfigurability at runtime using a set of software primitives \cite{AbadalACCESS}. To this end, \acp{SDM} require the integration of a (possibly wireless) network of controllers within the metamaterial, where each controller interacts locally and communicates globally to obtain the programmed behavior. Since the metamaterial building blocks are generally around $\lambda$/10 in size, \acp{SDM} for mmWave or terahertz applications may require the use of microscale antennas to interconnect the controllers. Unlike WNSNs, however, the \ac{SDM} network is highly integrated within a single physical device. This may complicate the antenna design and the channel characterization, as multiple sources of coupling and interference appear. Two examples of such impairments are near-field effects between adjacent antennas or the interference between the waves used to communicate and the waves that the metamaterial aims to manipulate.

\noindento {\bf Wireless Network-on-Chip (WNoC).} Multicore processors require an on-chip network to interconnect the \emph{multi}ple computing \emph{cores} that work in parallel. With the advent of manycore processors, the \ac{WNoC} paradigm advocates for placing multiple antennas within the same chip to wirelessly communicate its cores, thereby complementing existing wired networks to address their performance and efficiency bottlenecks \cite{AbadalMICRO}. In this scenario, usage of chip area is a primary design constraint and must be minimized, especially as more cores are integrated within the same chip. Yet more important are the high bandwidth requirements, in the 10--100 Gbps range, which together with the chip area constraints suggest the use of graphene-based terahertz antennas. As in the \ac{SDM} scenario, \acp{WNoC} are integrated in a planar and highly integrated environment, which challenges the antenna design and channel characterization. In this case, multipath and the interaction of electromagnetic waves with the chip package are important concerns.

\section{Conclusion}
\label{sec:conclusion}
Graphene-based antennas present unique miniaturization and tunability features in the terahertz band. These, however, will come at the cost of reduced radiation efficiency unless the quality of the graphene monolayers is improved in the near future. We have observed that stacks containing multiple graphene monolayers and thin dielectric layers exhibit similar tradeoffs between the different performance characteristics. This analysis may guide the design of antennas for area-constrained applications, as scenarios like \ac{SDM} may tolerate higher loss in exchange for miniaturization, whereas scenarios like \ac{WNoC} may give priority to performance and reconfigurability over miniaturization.




\begin{thebibliography}{10}
\providecommand{\url}[1]{#1}
\csname url@samestyle\endcsname
\providecommand{\newblock}{\relax}
\providecommand{\bibinfo}[2]{#2}
\providecommand{\BIBentrySTDinterwordspacing}{\spaceskip=0pt\relax}
\providecommand{\BIBentryALTinterwordstretchfactor}{4}
\providecommand{\BIBentryALTinterwordspacing}{\spaceskip=\fontdimen2\font plus
\BIBentryALTinterwordstretchfactor\fontdimen3\font minus
  \fontdimen4\font\relax}
\providecommand{\BIBforeignlanguage}[2]{{%
\expandafter\ifx\csname l@#1\endcsname\relax
\typeout{** WARNING: IEEEtran.bst: No hyphenation pattern has been}%
\typeout{** loaded for the language `#1'. Using the pattern for}%
\typeout{** the default language instead.}%
\else
\language=\csname l@#1\endcsname
\fi
#2}}
\providecommand{\BIBdecl}{\relax}
\BIBdecl

\bibitem{Schwierz2013}
F.~Schwierz, ``{Graphene Transistors: Status, Prospects, and Problems},''
  \emph{Proceedings of the IEEE}, vol. 101, no.~7, pp. 1567--1584, 2013.

\bibitem{Vakil2011}
A.~Vakil and N.~Engheta, ``{Transformation optics using graphene},''
  \emph{Science}, vol. 332, no. 6035, pp. 1291--4, 2011.

\bibitem{Bao2012}
Q.~Bao and K.~P. Loh, ``{Graphene photonics, plasmonics, and broadband
  optoelectronic devices},'' \emph{ACS Nano}, vol.~6, no.~5, pp. 3677--94,
  2012.

\bibitem{AbadalTCOM}
S.~Abadal, I.~Llatser, A.~Mestres, H.~Lee, E.~Alarc\'{o}n, and
  A.~Cabellos-Aparicio, ``{Time-domain analysis of graphene-based miniaturized
  antennas for ultra-short-range impulse radio communications},'' \emph{IEEE
  Transactions on Communications}, vol.~63, no.~4, pp. 1470--1482, 2015.

\bibitem{Geim2007}
A.~K. Geim and K.~S. Novoselov, ``{The rise of graphene},'' \emph{Nature
  materials}, vol.~6, no.~3, pp. 183--191, 2007.

\bibitem{Hanson2008a}
G.~W. Hanson, ``{Dyadic Green's functions and guided surface waves for a
  surface conductivity model of graphene},'' \emph{Journal of Applied Physics},
  vol. 103, no.~6, p. 064302, 2008.

\bibitem{Perruisseau-Carrier2013}
J.~Perruisseau-Carrier, M.~Tamagnone, J.~S. G\'{o}mez-D\'{i}az, and E.~Carrasco,
  ``{Graphene antennas: Can integration and reconfigurability compensate for
  the loss?}'' \emph{Proceedings of the EuMC '13}, 2013, pp. 369--372.

\bibitem{Yu2009Chemical}
Y.-J. Yu, Y.~Zhao, S.~Ryu, L.~E. Brus, K.~S. Kim, and P.~Kim, ``{Tuning the
  graphene work function by electric field effect},'' \emph{Nano Letters},
  vol.~9, no.~10, pp. 3430--3434, 2009.

\bibitem{CST}
\BIBentryALTinterwordspacing
``{CST Microwave Studio}.'' [Online]. Available: \url{http://www.cst.com}
\BIBentrySTDinterwordspacing

\bibitem{Cabellos2014}
A.~Cabellos-Aparicio, I.~Llatser, E.~Alarc{\'{o}}n, A.~Hsu, and T.~Palacios,
  ``{Use of THz photoconductive sources to characterize tunable graphene RF
  plasmonic antennas},'' \emph{IEEE Transactions on Nanotechnology}, vol.~14,
  no.~2, pp. 390--396, 2015.

\bibitem{Gomez2015}
J.~S. G{\'{o}}mez-D{\'{i}}az, C.~Moldovan, S.~Capdevila, J.~Romeu, L.~Bernard,
  A.~Magrez, A.~Ionescu, and J.~Perruisseau-Carrier, ``{Self-biased
  reconfigurable graphene stacks for terahertz plasmonics},'' \emph{Nature
  Communications}, vol.~6, no. 6334, 2015.

\bibitem{Hosseininejad2017}
S.~E. Hosseininejad, E.~Alarc{\'{o}}n, N.~Komjani, S.~Abadal, M.~C. Lemme,
  P.~Haring Bol{\'{i}}var, and A.~Cabellos-Aparicio, ``{Study of hybrid and pure 
	plasmonic terahertz antennas based on graphene guided-wave structures},'' 
	\emph{Nano Communication Networks}, vol.~12, pp. 34--42, 2017.

\bibitem{Jornet2014TRANSCEIVER}
J.~M. Jornet and I.~F. Akyildiz, ``{Graphene-based plasmonic nano-transceiver
  for terahertz band communication},'' in \emph{Proceedings of the EuCAP '14},
  2014, pp. 492--6.

\bibitem{Hosseininejad2016a}
S.~E. Hosseininejad and N.~Komjani, ``{Waveguide-fed tunable terahertz antenna
  based on hybrid graphene-metal structure},'' \emph{IEEE Transactions on
  Antennas and Propagation}, vol.~64, no.~9, pp. 3787--3793, 2016.

\bibitem{Huang2012ARRAY}
Y.~Huang, L.~Wu, M.~Tang, and J.~Mao, ``{Design of a beam reconfigurable THz
  antenna with graphene-based switchable high-impedance surface},'' \emph{IEEE
  Transactions on Nanotechnology}, vol.~11, no.~4, pp. 836--842, 2012.

\bibitem{Carrasco2013}
E.~Carrasco and J.~Perruisseau-Carrier, ``{Reflectarray antenna at terahertz
  using graphene},'' \emph{IEEE Antennas and Wireless Propagation Letters},
  vol.~12, pp. 253--256, 2013.

\bibitem{Akyildiz2010}
I.~F. Akyildiz and J.~M. Jornet, ``{The Internet of nano-things},'' \emph{IEEE
  Wireless Communications}, vol.~17, no.~6, pp. 58--63, 2010.

\bibitem{AbadalACCESS}
S.~Abadal, C.~Liaskos, A.~Tsioliaridou, S.~Ioannidis, A.~Pitsillides,
J.~Sol{\'{e}}-Pareta, E.~Alarc{\'{o}}n, and A.~Cabellos-Aparicio,
 ``{Computing and communications for the software-defined metamaterial paradigm: 
A context analysis},'' \emph{IEEE Access}, vol.~5, 2017.

\bibitem{AbadalMICRO}
S.~Abadal, B.~Sheinman, O.~Katz, O.~Markish, D.~Elad, Y.~Fournier, D.~Roca,
  M.~Hanzich, G.~Houzeaux, M.~Nemirovsky, E.~Alarc{\'{o}}n, and
  A.~Cabellos-Aparicio, ``{Broadcast-enabled massive multicore architectures: A
  wireless RF approach},'' \emph{IEEE MICRO}, vol.~35, no.~5, pp. 52--61, 2015.

\end{thebibliography}
%



\end{document}